\documentstyle[aps,prb,epsf]{revtex}
\draft
\begin{document}
\bibliographystyle{unsrt}
\title{
Quasiparticle structure and coherent propagation in
the $t-J_{z}-J_{\perp}$ model
}
\author{
Junwu Gan and Per Hedeg\aa{}rd\cite{Copenhagen}
}
\address{
 Department of Physics,
      University of California, Berkeley, CA 94720.
}

\maketitle

\begin{abstract}

 Numerical studies, from variational calculation to
exact diagonalization, all indicate
that the  quasiparticle generated by introducing one hole
into a two-dimensional quantum antiferromagnet has the same
nature as a string state in the  $t-J_{z}$ model.
Based on this observation,  we attempt to
visualize the quasiparticle formation and subsequent
coherent
propagation at low energy  by studying
the generalized $t-J_{z}-J_{\perp}$ model
in which we first diagonalize the $t-J_{z}$ model and then
perform a {\em degenerate}
perturbation  in $J_{\perp}$.
We construct
the quasiparticle state and derive an effective Hamiltonian
describing the coherent  propagation of the quasiparticle
 and its interaction with the
spin wave excitations in the presence
of the N\'{e}el order.
We expect that qualitative properties of the
quasiparticle remain intact
when analytically continuing $J_{\perp}$
from the anisotropic $J_{\perp} < J_{z}$ to the isotropic
$J_{\perp} = J_{z}$ limit, despite the fact that
the spin wave excitations change from gapful to gapless.
Extrapolating to $J_{\perp}=J_{z}$, our quasiparticle
dispersion and spectral weight compare
well with the exact numerical results for small clusters.

\end{abstract}

\vskip 30pt


%=================================================

			% BODY OF PAPER
%\begin{multicols}{2}

\section{Introduction}

Since the beginning of the era of high temperature
superconductivity,
it has been suggested that
the one-band Hubbard model or the
closely
related $t-J$ model
describing the CuO$_{2}$ plane may be the basic model
for cuprate superconductors.\cite{ande87,zhang88}
While the early suggestion was based on the high energy
properties of the cuprates,
such as electronic structure and
spectroscopy, it is remarkable that many peculiar
low energy properties of the cuprates
have been reproduced in exact numerical studies of the
Hubbard or $t-J$ model.\cite{dago94}
Thus, understanding the two-dimensional Hubbard or $t-J$
model
stands out as a central problem in the field of
high temperature superconductivity.

While the Hubbard and $t-J$ models
at half-filling describing
the
insulating parent compound
are well understood by now,\cite{chak89}
present
understanding of the effect of doping mostly derives
from numerical
studies.\cite{dago94,trug88,trug90,dago90,liu92,poil93,poil93_1}
Essentially, the numerical results can be summarized as
follows:
(i) Doping the two-dimensional (2D) antiferromagnet with one hole
generates a quasiparticle with finite spectral weight
which implies that the quasiparticle
carries charge $e$ and spin $1/2$.
(ii) The quasiparticle dispersion has a bandwidth of order
$J$, is almost flat near $(\pi,0)$ and $(0,\pi)$,
and the general structure agrees
with the recent photoemission results.
 (iii) The remaining spectral weight is
mostly distributed in
the energy range corresponding to the mid-infrared
absorption
universally observed in the optical experiments
of the high $T_{c}$ cuprates\cite{tanner92}.
(iv) There is tentative evidence indicating
that the quasiparticle band
found in the one-hole case remains fairly robust
under the finite
doping.\cite{eder94,haas94,moreo94,moreo94_1,preu}
The latter may have already received experimental
support.\cite{aebi94}
In this paper, we attempt to provide  an intuitive way
to understand all numerical results, and
derive a low-energy effective
Hamiltonian for the one-hole
problem. We shall also point out
the extension to finite doping
when the N\'{e}el order is destroyed.

The exact numerical results  on the one-hole problem in
small clusters
suggest an adiabatic picture for the doped antiferromagnet.
For the $t-J$ model with $t \gg J$, the hole hopping takes
place at
a typical time scale of $1/t$ while the time scale of spin
fluctuations
is at least $1/J$($\gg 1/t$).
When considering the hole hopping,
it is reasonable to take a snapshot of
the spin configuration and freeze it.
When the system possesses either long range or short
range antiferromagnetic correlation, a typical
spin configuration also possesses the same kind of
correlation.
As the hole hops in two dimensions, it leaves behind
 a string of overturned spins along its trace.
These overturned spins are energetically frustrated with
their
neighboring spins surrounding the trace.
The energy increase due to the frustration is roughly
proportional
to the length of the string.
Therefore, the hole is linearly
confined to its original
site.
Due to this confinement, the hole is dressed by a cloud
of local distortion of the  spin configuration.
The quasiparticle is just such a composite object.
The formation of the quasiparticle does not require the
existence of long range antiferromagnetic order.
The quasiparticle should be stable as long as its
size is smaller than the magnetic correlation length.
It is also important to realize that the formation of the
quasiparticle
is a high-energy  process at the
scale of the hopping amplitude $t$. Exact diagonalization
study of
small clusters provides accurate description of the
quasiparticle structure.
At the lower-energy scale of order $J$,
the physical properties of the
doped antiferromagnet are described by
the coherent propagation of the quasiparticle
and its interaction with
the spin wave excitations.
It is at this lower-energy scale that
an effective Hamiltonian is needed.

Our approach amounts to a
degenerate perturbative treatment
of the
transverse spin fluctuation term $J_{\perp}$
of the $t-J_{z}-J_{\perp}$ model.
We first construct a basis of states by diagonalizing
the $t-J_{z}$ model. Then we project the transverse
$J_{\perp}$
term onto the constructed basis of states and use the
resulting matrix elements as an effective representation of
the transverse spin fluctuation term.
Previous analytical
studies\cite{liu92,kane89,su,ruckenstein}
have concentrated on solving Bogoliubov-deGennes-type
equations (summing up the noncrossing diagrams) in which
the spin wave excitations are represented by
  Holstein-Primakoff bosons\cite{holstein40} and the hole
hopping term
is treated as the interaction.
Although the plausible argument for
the ``dominant-pole
approximation'' put forward by
Kane, Lee, and Read\cite{kane89}
indicated the existence of a low-energy quasiparticle,
it has been found that
a self-consistent solution is needed
to obtain the correct features
of the quasiparticle
dispersion and
spectral weight.
The self-consistent solution can be obtained only
numerically.\cite{liu92,ruckenstein}
The advantage of our approach is the clear separation
between the quasiparticle formation at high energy
from the low-energy coherent propagation of the
quasiparticle.
The parameters in the effective Hamiltonian describing the
quasiparticle
propagation are evaluated directly, though approximately.
Most importantly, the form of the effective Hamiltonian
does not depend on the detailed calculations and
is completely determined by the spin and charge
quantum numbers of the
quasiparticle.

The specific nature of the quasiparticle, that it is
essentially same as a string state
 as explicitly shown by Eder and Ohta,\cite{eder94p}
 implies that
the qualitative properties of the quasiparticle
remain intact
when analytically
continuing $J_{\perp}$ from the anisotropic
$J_{\perp}< J_{z}$
to the isotropic limit $J_{\perp}=J_{z}$,
despite the fact
that the spin wave excitations change from gapful to
gapless.
We have evaluated the parameters of the effective
Hamiltonian
to first order in $J_{\perp}$ which is appropriate
when $J_{\perp} \ll J_{z}$. Nevertheless,
by taking $J_{\perp} = J_{z}$ our quasiparticle dispersion
and spectral weight compare qualitatively and quantitatively
well with the exact diagonalization results
for the isotropic $t-J$ model.
The quantitative agreement is expected to improve
should  our results be compared with the exact numerical
results
for the anisotropic $t-J_{z}-J_{\perp}$ model if available.
These expectations can be critically tested.

Although the hole hopping amplitude $t$ is much larger
than $J$, the bare hole hopping is localized inside the
quasiparticle
and the associated excitations are gapful.
The true low-energy charge excitations are those
corresponding
to the coherent propagation of the quasiparticle.
The velocity of the quasiparticle near the bottom of the
quasiparticle band is
much smaller than the spin wave velocity.
Therefore, the spin wave excitation is not
dissolved into the quasiparticle-hole continuum
excitations. As a result,
the low-energy spin wave excitations cannot decay
by exciting the quasiparticle into a higher-energy
state of the quasiparticle band.
The linear dispersion of the spin excitations is
preserved,\cite{sachdev92,chubukov93,sokol94}
 contrary to naive expectation based on the fact
$t \gg J$.

The layout of the paper is as follows.
In Sec.\ref{sec2}, we derive the effective Hamiltonian for
one hole
in the antiferromagnet.
In Sec.\ref{sec3}, we compare our quasiparticle dispersion
and spectral
weight with the exact diagonalization results in small
clusters.
We conclude the paper in Sec.~\ref{sec4}.

\section{Adiabatic description of a doped antiferromagnet}
\label{sec2}

In this section, we shall derive an effective Hamiltonian
describing the
quasiparticle motion in a doped antiferromagnet and
the interaction of the quasiparticle
with long wavelength spin wave excitations in the presence
of N\'{e}el order.
 We shall first construct
all low-energy eigenstates of
the $t-J_{z}$ Hamiltonian. Then we represent the transverse
Heisenberg
interaction $J_{\perp}$ term
by its matrix elements
within the basis of these eigenstates of the $t-J_{z}$
Hamiltonian.

\subsection{Construction of the basis states}

The $t-J$ model Hamiltonian has the form
\begin{eqnarray}
H &=& H_{0} + H_{1}  , \\
H_{0} &=& -t \sum_{<i,j>}
[ (1-n_{i,-\sigma})c_{i\sigma}^{\dagger} c_{j\sigma}(1-
n_{j,-\sigma}) + h.c.]
+ J_{z} \sum_{<i,j>} S_{i}^{z} S_{j}^{z} , \\
H_{1} &=& J_{\perp} \sum_{<i,j>}
( S_{i}^{x} S^{x}_{j} + S_{i}^{y} S_{j}^{y} ) ,  \label{h1}
\end{eqnarray}
where
$c^{\dagger}_{i\sigma}$ creates an electron at lattice site
$i$
with spin $\sigma$,
 $n_{i,\sigma}=c^{\dagger}_{i\sigma} c_{i\sigma}$
is the electron density operator,
and $\vec{S}_{i} = \sum_{\mu\nu} c^{\dagger}_{i\mu}
\vec{\sigma}_{\mu\nu} c_{i\nu} /2 $ is the spin operator.
The summation $< \! i, j \!>$ is limited to the nearest
neighbor links.
Unlike the N\'{e}el state in the undoped case,
the ground state
of the $t-J_{z}$ Hamiltonian $H_{0}$ in
the presence of one
hole has  macroscopic near degeneracy.
These ground states can be generated
by removing a spin at an arbitrary site in the N\'{e}el
state and
allowing the hole to hop. As the hole hops, it leaves
behind a string
of overturned spins. Therefore, the hole suffers a linear
binding
potential to its original site. The eigenstates of the hole
consist of discrete bound states with energy spacing
uniquely characterized by  an energy scale
$ ( J_{z}/t)^{2/3}t$.
For the purpose of deriving a low-energy effective theory,
we only
need to retain the ground state.

There is an excellent approximation
to obtain  the ground  states of
the $t-J_{z}$ Hamiltonian.\cite{brink70,shrai88}
The approximation amounts to neglecting winding paths (Bethe
lattice approximation).
We briefly describe this approximation here
with some minor improvement. We shall also use slightly more
concrete notations to meet our needs. The perfect
classical  N\'{e}el state
with only one hole can be labeled by the position of the
hole,
${\bf r}$. There are no frustrated spins in this state.
Applying the hole hopping operator in $H_{0}$
to this state generates four new
configurations, each with a string of a length
of one lattice spacing,
$| {\bf r}, \{ {\bf a } \} \rangle $, where ${\bf a}$ can be
one of the four unit lattice vectors $\pm {\bf x}$ or
$\pm {\bf y}$,  representing the hopping
of the hole from ${\bf r}$ to
${\bf r}+{\bf a}$.
 From the four configurations we construct
an ``$s$-wave'' state with string length 1,
  \begin{equation}
	| {\bf r}, 1 \rangle = \frac{1}{2}
\sum_{{\bf a} = \pm {\bf x}, \pm {\bf y}}
	| {\bf r}, \{ {\bf a }  \} \rangle .
\label{string1}
  \end{equation}
Other states with longer strings are constructed similarly,
  \begin{equation}
	| {\bf r}, l \rangle
 = \frac{1}{2 \times 3^{(l-1)/2} }
	\sum_{ {\bf a}_{1} \cdots {\bf a}_{l} }
	| {\bf r}, \{ {\bf a}_{1}, {\bf a}_{2}, \cdots
		{\bf a}_{l} \} \rangle ,
\label{stringl}
  \end{equation}
where the summation is implicitly subject to the constraint
${\bf a}_{i} \neq -{\bf a}_{i-1}$
to prevent hole retracing, here and  throughout the paper.
The Bethe lattice approximation is to use the matrix
elements within the set
of  states $|{\bf r},l\rangle$ to represent the $t-J_{z}$
Hamiltonian
for the corresponding subspace,
\begin{eqnarray}
H_{0} |{\bf r},0 \rangle &=& J_{z} | {\bf r},0 \rangle - 2
t \, |{\bf r}, 1\rangle ,
		\nonumber  \\
H_{0} |{\bf r}, 1 \rangle &=& \frac{5}{2} J_{z}
|{\bf r},1\rangle
	-2 t \, | {\bf r},0 \rangle
	- \sqrt{3} \, t \, |{\bf r}, 2\rangle ,
		    \label{tjzl}           \\
H_{0} |{\bf r}, l\rangle & \simeq & \epsilon_{l} |{\bf r},
l\rangle
 -\sqrt{3} \, t \,
\left( |{\bf r}, l-1\rangle + | {\bf r}, l+1 \rangle
\right) ,
	\hspace{.2in} l \geq 2 ,  \nonumber
\end{eqnarray}
where
\begin{equation}
\epsilon_{l} = J_{z} \langle l, {\bf r} |  \sum_{<i,j>}
\left( S_{i}^{z} S_{j}^{z}
+ \frac{1}{4} \right) | {\bf r}, l\rangle .
\end{equation}
The usage of the average energy $\epsilon_{l}$ slightly
improves the
approximation~(\ref{tjzl}) for the $t-J_{z}$ Hamiltonian
used by Shraiman and Siggia.\cite{shrai88}
In this paper, we shall directly evaluate
the first few values of $\epsilon_{l}$
up to $l=7$. For larger $l$ values,
we shall use the extrapolation of  $\epsilon_{l}$
from small $l$.
The values of $\epsilon_{l}$ used in this paper are listed
in Table~\ref{elvalue}.
The extrapolation is
$\epsilon_{l} = 2.574 - 0.0156 (-1)^{l} - 0.7857/l
+ 0.6857 l $.
Numerically diagonalizing (\ref{tjzl})
determines the coefficients $u_{l}$ in  the approximate
ground
state of the $t-J_{z}$ Hamiltonian,
\begin{equation}
H_{0} |{\bf r}\rangle = - E_{0} |{\bf r}\rangle,
\hspace{.2in} |{\bf r}\rangle =
\sum_{l=0}^{\infty} u_{l} \, |{\bf r}, l\rangle .
		\label{e0r}
\end{equation}
There is
a macroscopic number of such ground
states since we can choose
${\bf r}$ to be any
lattice site.
The size of
$ |{\bf r}\rangle $
depends on $J_{z}/t$ ratio:
The smaller the ratio, the bigger the
size.

In the two-dimensional
square lattice, the macroscopic degeneracy
of $ |{\bf r}\rangle $ is removed
by the winding
paths (so-called Trugman process).\cite{trug88}
This is reflected in the fact that
the approximate ground
states $|{\bf r}\rangle$ labeled by
${\bf r}$ and given by Eq.~(\ref{e0r}) are
not strictly orthogonal to each other.
In the presence of N\'{e}el order, we can divide the
two-dimensional
square lattice into
 ${\cal A}$ and ${\cal B}$ sublattices.
For ${\bf r}$ belonging to the same sublattice, there is a
small overlap.
An example is shown in Fig.~\ref{nonorthog}.
This configuration belongs to both
$|{\bf r}={\bf r}_{0}, l=5\rangle$
and $|{\bf r}={\bf r}_{2}, l=1\rangle$ according to the
prescription (\ref{stringl}),
causing an overlap between $|{\bf r}_{0}\rangle$
and $|{\bf r}_{2}\rangle$ states.
For $J_{z}/t=0.3$, the overlap between
$|{\bf r} \rangle$  and
$|{\bf r} + {\bf x} + {\bf y} \rangle$
is $0.0409$.
The overlap between $|{\bf r} \rangle$  and
$|{\bf r} + 2{\bf x}  \rangle$
is 0.0019, negligible even compared
to $\langle {\bf r}|{\bf r}+{\bf x}+{\bf y} \rangle$.
For  ${\bf r}$
belonging to different sublattices, any
two states are orthogonal since they
have different  $S^{z}_{tot}=\sum_{i} S_{i}^{z}$.
The small corrections due to the
nonorthogonality of $|{\bf r} \rangle$
will be taken into account
later by using tight-binding approximation
to study the propagation of
$|{\bf r} \rangle$ as a whole.

The set of states $|{\bf r}\rangle$ is far from complete.
One way to complement the set of states $|{\bf r}\rangle$ is
to include
 ``non-$s$-wave'' string  states.
Instead (\ref{string1}), for instance,
 we can  form the $d$ state,
  \begin{equation}
	| {\bf r}, 1 \rangle_{d} = \frac{1}{2} \left(
	| {\bf r}, \{ {\bf x }  \}
\rangle -  | {\bf r}, \{ {\bf y }  \} \rangle
+ | {\bf r}, \{ -{\bf x }  \} \rangle -  | {\bf r}, \{ -{\bf
y }  \} \rangle \right).
	\label{string1p}
  \end{equation}
Applying the hole hopping operator in $H_{0}$
to (\ref{string1p}) will generate
a new set of states.
Diagonalizing the $t-J_{z}$ Hamiltonian
within the corresponding new subspace
will give us new eigenstates of the $t-J_{z}$
Hamiltonian whose energies are about
 $1.5 J_{z}$ higher than the states $|{\bf r}\rangle$
given by Eq.~(\ref{e0r}).
Obviously, the new eigenstates are also localized
around ${\bf r}$.
For the same reason --- the existence of the winding paths
--- the new eigenstates centered at ${\bf r}$
are not strictly orthogonal to the ``$s$-wave''
states $|{\bf r}'\rangle$ centered
at a different site ${\bf r}'$ in
the same sublattice
and not too far away from ${\bf r}$.
We note that if ${\bf r}={\bf r}'$ there is no overlap
due to the different rotational symmetries.
The spin fluctuation term (\ref{h1}) also
has nonvanishing matrix elements
between the ``$s$-wave'' states $|{\bf r}'\rangle$
and the new eigenstates.
However, both overlaps and matrix elements are very
small.
Thus,  these new eigenstates of the $t-J_{z}$ Hamiltonian
generated from ``non-$s$-wave'' string configurations
are almost disconnected from the subspace where the states
$|{\bf r}\rangle$
``live''. The ``non-$s$-wave'' states  will be omitted in
our low-energy effective theory.

The eigenstates of the $t-J_{z}$ Hamiltonian that
we have discussed so
far contain overturned spins only along the trace of the
hole.
If we neglect the winding paths,
the overturned spins form a continuous string in each
spin configuration.
We need to complement the set of states $|{\bf r}\rangle$ by
including one or more isolated overturned
spins outside the string.
Inside the eigenstates $|{\bf r}\rangle$, the hole is
localized around
the site ${\bf r}$, and so are all the strings of overturned
spins.
Thus, if the  isolated overturned spins are far from
${\bf r}$, they will not
interfere with the strings.
In this case, we have simple superposition of the isolated
overturned spins and the string state $|{\bf r}\rangle$.
 More delicate situations occur when the isolated
overturned spins
are located close to the center site ${\bf r}$
of the string state.
Our goal is to find all new eigenstates of the $t-J_{z}$
Hamiltonian
that can be connected to the
string states $|{\bf r}\rangle$ by applying $H_{1}$.

We start from a configuration with one
 isolated overturned  spin
at ${\bf r}_{s}$ and a hole at ${\bf r}$
in an otherwise perfect N\'{e}el background.
We demand that a new subspace be generated
by applying the hole hopping operator to
this configuration
and new eigenstates of the $t-J_{z}$ Hamiltonian
be generated by
diagonalizing the $t-J_{z}$ Hamiltonian
 within this new subspace.
In order for $H_{1}$ to have
nonvanishing matrix elements
between the new states and the old states given by
Eq.~(\ref{e0r}),
the site of the hole ${\bf r}$
and the site of the overturned spin
${\bf r}_{s}$ must belong to different sublattices.
Otherwise, the generated new state has $S^{z}_{tot}=\pm 3/2
$ and cannot
be connected to the states (\ref{e0r})
by $H_{1}$ since the states (\ref{e0r})
have $S^{z}_{tot}=\pm 1/2$.
Since we require ${\bf r}_{s}$ to be close to ${\bf r}$,
the first possibility
 is for ${\bf r}_{s}$ to be a nearest neighbor
of ${\bf r}$. But this is a string of length 1
 $|{\bf r}_{s}, \{ {\bf r}-{\bf r}_{s} \} \rangle$
which we have already considered.
Thus, no new subspace will be generated from this
configuration.
The next possibility is for ${\bf r}_{s}$ to be farther away
from
 ${\bf r}$: ${\bf r}_{s} - {\bf r} = 2 {\bf x} + {\bf y}$
or ${\bf r}_{s} -{\bf r} = 3 {\bf x} $.
Letting the hole hop in these
configurations  will generate new subspaces.
In principle, we have to diagonalize the $t-J_{z}$
Hamiltonian
within these new subspaces and keep the low-energy
eigenstates.
In practice, we shall do it approximately as in (\ref{e0r}).
In complete analogy with (\ref{stringl}),
we  construct generalized states
	\begin{equation}
| {\bf r}_{s}, {\bf r}, l \rangle =  \frac{1}{2 \times
3^{(l-1)/2} }
	\sum_{ {\bf a}_{1} \cdots {\bf a}_{l} }
	| {\bf r}_{s}, {\bf r}, \{ {\bf a}_{1}, {\bf
a}_{2}, \cdots
		{\bf a}_{l} \} \rangle .
\label{rsrl}
\end{equation}
Then we use the matrix elements of the $t-J_{z}$
Hamiltonian within the set of states $ | {\bf r}_{s}, {\bf
r}, l \rangle$
for $l=0,1,\cdots \infty$ as an approximate representation
of the $t-J_{z}$ Hamiltonian. This representation has
exactly the same form
as (\ref{tjzl}) except for different diagonal energies
$\epsilon'_{l}$:
\begin{eqnarray}
H_{0} |{\bf r}_{s}, {\bf r},0 \rangle
 &=& 3 J_{z} | {\bf r}_{s}, {\bf r},0 \rangle
	- 2  t \, |{\bf r}_{s},{\bf r}, 1\rangle ,
			 \nonumber  \\
H_{0} |{\bf r}_{s},{\bf r}, 1 \rangle &=&
	\frac{9}{2} J_{z} |{\bf r}_{s},{\bf r},1\rangle
	-2 t \, | {\bf r}_{s}, {\bf r},0 \rangle
	- \sqrt{3} \, t \, |{\bf r}_{s}, {\bf r},
2\rangle ,
			\label{tjzlh}          \\
H_{0} |{\bf r}_{s}, {\bf r}, l\rangle & \simeq &
	\epsilon'_{l} |{\bf r}_{s}, {\bf r}, l\rangle
 -\sqrt{3} \, t \, \left( |{\bf r}_{s},{\bf r}, l-1\rangle
 + | {\bf r}_{s}, {\bf r}, l+1 \rangle  \right) ,
	\hspace{.2in} l \geq 2 ,  \nonumber
\end{eqnarray}
where
\begin{equation}
\epsilon'_{l} = J_{z} \langle l,{\bf r},{\bf r}_{s}|
 \sum_{<i,j>} \left( S_{i}^{z} S_{j}^{z} + \frac{1}{4}
\right)
|{\bf r}_{s}, {\bf r}, l\rangle .
\end{equation}
The values of $\epsilon'_{l}$ used in this paper are  also
listed
in Table~\ref{elvalue}.
For large $l$, we use
extrapolated values from $\epsilon'_{l}$
listed in Table~\ref{elvalue}
in the same way as we did for $\epsilon_{l}$.
Diagonalizing (\ref{tjzlh}) gives us
new approximate   eigenstates
\begin{eqnarray}
H_{0} \, |{\bf r}_{s},{\bf r}\rangle &=& - E'_{0}({\bf
r}_{s}-{\bf r}) \,
|{\bf r}_{s},{\bf r}\rangle,          \nonumber \\
 |{\bf r}_{s},{\bf r}\rangle &=& \sum_{l=0}^{\infty}
 u'_{l}({\bf r}_{s}-{\bf r}) \, |{\bf r}_{s}, {\bf r},
l\rangle .
		\label{e0rs}
\end{eqnarray}

Following the same line of reasoning,
we can construct further approximate
eigenstates of the $t-J_{z}$ Hamiltonian.
A straightforward extension is to consider ${\bf r}_{s}$
being  further away from ${\bf r}$, for instance,
${\bf r}_{s}-{\bf r}=5{\bf x}, 4{\bf x}+{\bf y}$, or $3{\bf
x}+2{\bf y}$.
Since  in the realistic situation  $J/t = 0.2 \sim 0.4$,
the size of the string state $|{\bf r}\rangle$ is limited
within
$l \leq 5$ inside (\ref{e0r}). Therefore, we can
approximately
treat the isolated overturned spin at ${\bf r}_{s}$ as
independent
of ${\bf r}$ if their separation is five or more
lattice spacings.
We should also try to generate new eigenstates by applying
the hole hopping operator in $H_{0}$ to
configurations with one hole and
an isolated pair of neighboring overturned spins
in an otherwise perfect N\'{e}el background.  If the hole is
a nearest neighbor of the pair of overturned spins,
this is a string of length 2
that we have already considered in (\ref{stringl}).
As the next  possibility, the hole could be separated
from
the pair of overturned spins by one lattice site.
One such example is shown in
the upper panel of Fig.~\ref{holeflip2}.
However, this configuration again has already appeared in
(\ref{rsrl})
in the  $l=3$ state as illustrated
in the lower panel of  Fig.~\ref{holeflip2}.
No new eigenstates
of the $t-J_{z}$ Hamiltonian can be
generated from this configuration.
Thus, we only need to
consider configurations in which the hole
is separated from the pair of overturned spins by two or
more
lattice sites. In this paper we shall neglect the
correlation
 between the hole hopping and the pair of overturned spins
in this kind of configurations.
Later we shall see that
what we have neglected are
the  interaction vertices
with four operators, two
quasiparticle and two magnon
operators.

\subsection{Derivation of the effective Hamiltonian}

To obtain the low-energy effective Hamiltonian, we
identify the many-body state $|{\bf r}\rangle$ with a
quasiparticle
located at site
${\bf r}$. It is obvious that the quasiparticle
carries
charge $e$ and spin $1/2$ as a hole
because there is local deficiency
of both charge and spin.
Furthermore, the quasiparticle spin points in the
opposite directions for ${\bf r}$ belonging to different
sublattices.
Thus we introduce a fermionic operator $f^{\dagger}_{{\bf
r},\sigma}$
to create the state $|{\bf r}\rangle$ from the N\'{e}el
background.  In the N\'{e}el state, we divide
the square lattice into ${\cal A}$ and ${\cal B}$
sublattices and assume that
the spins in the ${\cal A}$ sublattice
point in the up direction.
If ${\bf r} \in {\cal A}$, then the quasiparticle
spin is down.
We shall approximate
$ |{\bf r}_{s},{\bf r}\rangle \simeq S^{+}_{{\bf r}_{s}}
f^{\dagger}_{{\bf r},\downarrow }
| \mbox{ N\'{e}el} \rangle $ for
${\bf r} \in {\cal A}$ and
${\bf r}_{s} \in {\cal B}$.
Using the Holstein-Primakoff representation of the
spin operator,\cite{holstein40}
$S^{-}_{i} \simeq b^{\dagger}_{i,1}$ for $i \in {\cal A}$
and $S^{+}_{i} \simeq b^{\dagger}_{i,2}$
for $i \in {\cal B}$,
we can write
\begin{equation}
|{\bf r}_{s},{\bf r}\rangle \simeq b_{{\bf
r}_{s},2}^{\dagger}
f^{\dagger}_{{\bf r},\downarrow }
| \mbox{N\'{e}el} \rangle
	\hspace{.3in} {\rm for} \; \;
 {\bf r} \in {\cal A}, \;
{\bf r}_{s} \in {\cal B}.
		\label{r_rs}
\end{equation}
The expression for ${\bf r} \in {\cal B}$ and
${\bf r}_{s} \in {\cal A} $ is similar.
The correlation between the hole hopping and the
overturned spin at ${\bf r}_{s}$ lowers the energy of the
state
$ |{\bf r}_{s},{\bf r}\rangle $ with respect to that of two
independent $b_{{\bf r}_{s}}^{\dagger}$ and
$ f^{\dagger}_{{\bf r},\sigma} $.
This energy difference amounts to an interaction vertex
 of the type
$f^{\dagger}_{{\bf r},\sigma}  f_{{\bf r},\sigma}
b_{{\bf r}_{s}}^{\dagger}  b_{{\bf r}_{s}} $.
In this paper, we shall
neglect all interaction vertices
with four operators
and assume that their effects can be properly accounted for
by small renormalization of
the parameters in the effective Hamiltonian.
This is supported by the
fact that quantum fluctuations only generate a small density
of overturned spins in the ground state of the 2D Heisenberg
antiferromagnet.

We have pointed out that
the essential difference from the undoped case lies
in the macroscopic near degeneracy of the
ground states
of the $t-J_{z}$ Hamiltonian.
The transverse spin fluctuation  $J_{\perp}$ term in the
$t-J_{z}-J_{\perp}$ Hamiltonian (\ref{h1}) lifts this
degeneracy
and generates  quasiparticle dispersion. Therefore,
only degenerate perturbation in $J_{\perp}$ is permitted
even for $J_{\perp} \ll J_{z}$.
In the following we shall represent the
transverse term (\ref{h1})
by its matrix elements
within the basis of the eigenstates of the $t-J_{z}$
Hamiltonian
that we have constructed in the previous subsection.
A typical example illustrating how the spin fluctuations
generate  quasiparticle hopping is shown in Fig.~\ref{hop1}.
Exchanging the first two spins on a string
cuts off its length
by two lattice units. Since the location of the
quasiparticle is
determined by the common starting point of the strings
inside
(\ref{stringl}) and (\ref{e0r}), this process moves
the quasiparticle to a nearby
site in the same sublattice.
Similarly, exchanging two spins near the starting point
but outside a string
increases the string length by two lattice units
and also generates  quasiparticle hopping.
We note that a quasiparticle cannot hop to the other
sublattice
simply because the quasiparticle spin directions are
opposite in
different sublattices.
To obtain the effective hopping amplitude of the quasiparticle,
we construct the propagating quasiparticle state
\begin{equation}
| {\bf k} \rangle = \frac{1}{\sqrt{N_{\bf k}}}
\sum_{{\bf r} \in {\cal A} } \exp(i {\bf k} \cdot {\bf r})
| {\bf r} \rangle ,		\label{qmove_k}
\end{equation}
where $1/\sqrt{N_{\bf k}}$ is the normalization factor.
Then we
calculate the quasiparticle  dispersion
according to the tight-binding approximation,\cite{callaway}
\begin{equation}
\epsilon_{\bf k} = \frac{1}{N_{\bf k}}
 \langle {\bf k} | H | {\bf k} \rangle ,
\end{equation}
where $H$ is the full $t-J$ Hamiltonian.
Using the fact that the string state
$| {\bf r} \rangle$ is an exact eigenstate of the
Hamiltonian
\begin{equation}
H_{r} = -t \sum_{<i,j>}
[ (1-n_{i,-\sigma})c_{i\sigma}^{\dagger} c_{j\sigma}(1-
n_{j,-\sigma}) + h.c.]
+ \sum_{l} \epsilon_{l} | {\bf r}, l \rangle \langle {\bf r},l | ,
\end{equation}
we can rewrite
\begin{equation}
\epsilon_{\bf k} = {\rm const} +
\frac{1}{N_{\bf k}}  \sum_{{\bf R} \in {\cal A}}
\exp(i {\bf k} {\bf R})
 \langle {\bf r} + {\bf R} |
H - H_{r}
| {\bf r} \rangle .		\label{epsk}
\end{equation}
The normalization factor
can be expanded in a power series,
\begin{equation}
\frac{1}{ N_{\bf k} } \simeq 1- 4
\langle {\bf r} +{\bf x}+{\bf y} |
{\bf r} \rangle
\cos k_{x} \cos k_{y} .  \label{exp_norm}
\end{equation}
For $J_{z}/t=0.2, \; 0.3, \; 0.4, \;
0.5 $, and  $ 0.6$, we have found
$ \langle {\bf r} +{\bf x}+{\bf y} | {\bf r} \rangle
= 0.065, \; 0.041, \; 0.027, \; 0.018$, and $ 0.012$
respectively.
The direct overlap
$\langle {\bf r}  + 2{\bf x} | {\bf r} \rangle$
is neglected
in (\ref{exp_norm}) because it is
too small even compared to
$\langle {\bf r} +{\bf x}+{\bf y} |
{\bf r} \rangle $.
Similarly, the matrix element
$ \langle {\bf r}+2{\bf x} | \left(
J_{z} \sum_{<i,j>} S_{i}^{z} S_{j}^{z}
- \sum_{l} \epsilon_{l} | {\bf r},
l \rangle \langle {\bf r},l | \right)
| {\bf r} \rangle $,
which appears in the calculation of
Eq.~(\ref{epsk}), can also be neglected
because the nonvanishing contributions
to this matrix element have the same origin
as in the direct overlap $\langle {\bf r}  + 2{\bf x} |
{\bf r}\rangle $.
After some
straightforward algebra, we obtain
from Eq.~(\ref{epsk})
\begin{equation}
\epsilon_{\bf k}
= {\rm const} + 4 \alpha_{1} \cos k_{x} \cos k_{y}
+ 2 \alpha_{2} ( \cos 2 k_{x} + \cos 2 k_{y} )
+ {\cal O}( \cos k_{x} \cos 3 k_{y}, \;
\cos 2k_{x} \cos 2 k_{y} ) ,   \label{epsk_1}
\end{equation}
where the high order trigonometric functions
corresponding to longer range quasiparticle hopping
are neglected. The coefficients in the dispersion are
given by
\begin{eqnarray}
\alpha_{1} &=&
 \langle {\bf r} +{\bf x}+{\bf y} | H-H_{r}
|{\bf r} \rangle
- \left[ \langle {\bf r}| H_{1}
|{\bf r} \rangle
+ 2 \langle {\bf r} + 2 {\bf x} | H_{1} | {\bf r} \rangle
\right]
\langle {\bf r} +{\bf x}+{\bf y} |{\bf r} \rangle ,
 	   \label{alpha1} \\
\alpha_{2} &=&
\langle {\bf r} + 2 {\bf x} | H_{1} | {\bf r} \rangle
- 2 \langle {\bf r} +{\bf x}+{\bf y} | {\bf r} \rangle
\langle {\bf r} +{\bf x}+{\bf y} | H-H_{r}
|{\bf r} \rangle .
 	  \label{alpha2}
\end{eqnarray}
This quasiparticle
dispersion can be described by a
tight-binding
Hamiltonian
\begin{equation}
H_{f} =  \frac{ 1 }{2}
\sum_{ {\bf r} \in {\cal A },\sigma=\pm}
\left[ \alpha_{1}
\sum_{{\bf a}=\pm{\bf x}\pm{\bf y}}
+ \, \alpha_{2}  \sum_{{\bf a}=\pm 2{\bf x},\pm 2{\bf y}}
\right]
\left( f^{\dagger}_{{\bf r},\sigma}
f_{{\bf r}+{\bf a},\sigma} + h.c. \right)  ,
\end{equation}
where ${\bf r} \in {\cal A}$ denotes summation over only one
sublattice. The summation over ${\cal A}$ and ${\cal B}$
sublattices is converted into the spin index $\sigma$ summation.

To obtain the coefficients $\alpha_{1}$ and
$\alpha_{2}$, we  need to calculate
the matrix elements of the type
$ \langle {\bf r}_{2} | \hat{O} | {\bf r}_{1} \rangle $
for several different operators $\hat{O}$.
We employ the following approximation when calculating
these matrix elements:
$ \langle {\bf r}_{2} | \hat{O} | {\bf r}_{1} \rangle
= \sum_{l_{1}+l_{2} \leq 10} u_{l_{1}} u_{l_{2}}
\langle {\bf r}_{2},l_{2} | \hat{O}
| {\bf r}_{1} , l_{1} \rangle $.
Each matrix element in the sum is then computed
straightforwardly.
By investigating how fast the eigenvector
$u_{l}$ decreases as a function of the string length,
one can be convinced that the approximation
$l_{1}+l_{2} \leq 10$ is sufficient for $J_{z} /t \geq 0.2$.
The values of $\alpha_{1}$ and $\alpha_{2}$
 for different $J_{z}/t$ ratios are listed in
Table~\ref{hop_vert}.

The spin fluctuation term $H_{1}$
exchanges two neighboring spins.
For most cases, this pair of
exchanged spins is far away from
the hole. If so,  the situation is
completely same as in the
undoped case.
Thus,  the Hamiltonian describing
 long wavelength spin excitations using the
Holstein-Primakoff
bosons has the familiar form,\cite{holstein40}
 upon neglecting the spin wave
interactions,
\begin{equation}
H_{b} = 2J_{z} \left[ \sum_{i \in {\cal A}}
b^{\dagger}_{i,1} b_{i,1}
+  \sum_{i \in {\cal B}} b^{\dagger}_{i,2} b_{i,2}  \right]
+ \frac{J_{\perp}}{2} \sum_{i \in {\cal A} }
\sum_{  {\bf a} = \pm {\bf x}, \pm {\bf y} }
( b^{\dagger}_{i,1} b^{\dagger}_{i+{\bf a}, 2} +
b_{i,1} b_{i+{\bf a}, 2}  ) .                   \label{hb}
\end{equation}
In principle, the summation over the nearest neighbors
in (\ref{hb}) should avoid a small region surrounding the
hole
to prevent double counting. However,
this restriction only results in interaction vertices
involving
four operators of the type $f^{\dagger} f b^{\dagger} b$ or
$f^{\dagger} f b^{\dagger} b^{\dagger}$.
These interactions are neglected in this paper.
Physically, these interactions occur because the
quasiparticle
has a finite spatial extent of a few lattice spacing.
But for long wavelength spin excitations the quasiparticle
can be viewed
as a point particle.

The spin fluctuation term $H_{1}$ also connects
the states given by Eqs.~(\ref{e0r}) and (\ref{e0rs}).
A typical example is shown in Fig.~\ref{vertex}.
This process allows
the quasiparticle to
jump over two sites, leaving
behind an isolated overturned spin.
We directly evaluate the following two overlap
coefficients:
\begin{eqnarray}
\lambda_{1}(J_{z}/t) &=& \frac{1}{J_{\perp}}
\langle {\bf r}| H_{1}
 |{\bf r},{\bf r}+2{\bf x}+{\bf y}\rangle  ,  \\
\lambda_{2}(J_{z}/t) &=& \frac{1}{J_{\perp}}
\langle {\bf r}| H_{1}
|{\bf r},{\bf r}+3{\bf x}\rangle  .
\end{eqnarray}
The nonorthogonality of the
approximate eigenstates
$| {\bf r} \rangle$
has a negligible effect
on the coefficients $\lambda_{1}$ and $\lambda_{2}$.
The results are also listed in Table~\ref{hop_vert}.
The corresponding part of the effective Hamiltonian is,
using Eq.~(\ref{r_rs}),
\begin{eqnarray}
H_{fb} &=&  J_{\perp}
  \sum_{ {\bf r} \in {\cal A } }
\left[ \lambda_{1}
\sum_{{\bf a}=\pm 2{\bf x}\pm{\bf y},
 \pm{\bf x}\pm 2{\bf y}}
+ \, \lambda_{2}  \sum_{{\bf a}=\pm 3{\bf x},\pm 3{\bf y}}
\right]
\left( f^{\dagger}_{{\bf r}+{\bf a},\uparrow}
f_{{\bf r},\downarrow} b^{\dagger}_{{\bf r},1} + h.c.
\right)
		  \nonumber \\
&+&   J_{\perp}
  \sum_{ {\bf r} \in {\cal B} }
\left[ \lambda_{1}
\sum_{{\bf a}=\pm 2{\bf x}\pm{\bf y},
\pm{\bf x}\pm 2{\bf y}}
+ \, \lambda_{2}  \sum_{{\bf a}=\pm 3{\bf x},\pm 3{\bf y}}
\right]
\left( f^{\dagger}_{{\bf r}+{\bf a},\downarrow}
f_{{\bf r},\uparrow}
b^{\dagger}_{{\bf r},2} + h.c. \right) .
\end{eqnarray}

The complete effective Hamiltonian is, upon
neglecting magnon interactions and
quasiparticle-magnon interactions containing four or more
operators,
\begin{equation}
H_{\rm eff} = H_{f} + H_{b} + H_{fb} .
\end{equation}
Since inside $H_{\rm eff}$ the lattice site
summation only extends within
one sublattice, the corresponding momentum summation should
be
limited within the
antiferromagnetic Brillouin zone(AFBZ).
This is due to the presence of the N\'{e}el order
which breaks the symmetry between the two sublattices.
In the momentum space, the three parts of $H_{\rm eff}$
have the following forms:
\begin{eqnarray}
H_{f} &=&\sum_{{\bf k} \in {\rm AFBZ}, \sigma=\pm}
\epsilon_{\bf k} \;
f^{\dagger}_{{\bf k},\sigma} f_{{\bf k},\sigma}    ,     \\
H_{b} &=&  \sum_{{\bf q} \in {\rm AFBZ}, }
 \left[  2 J_{z} ( b^{\dagger}_{{\bf q},1} b_{{\bf q},1}
+ b^{\dagger}_{{\bf q},2} b_{{\bf q},2} )
+ J_{\perp} \left( \cos q_{x} + \cos q_{y} \right)  \left(
b^{\dagger}_{{\bf q},1} b^{\dagger}_{-{\bf q},2}
+ b_{{\bf q},1} b_{-{\bf q},2}  \right)    \right]   ,
\\
H_{fb} &=&  \frac{1}{\sqrt{\cal N}} \sum_{{\bf k},{\bf q}
\in {\rm AFBZ} }
  \lambda_{\bf k} \left[
f^{\dagger}_{{\bf k},\uparrow}
f_{{\bf k}+{\bf q}, \downarrow} b^{\dagger}_{{\bf q},1}
+ f^{\dagger}_{{\bf k},\downarrow}
f_{{\bf k}+{\bf q}, \uparrow} b^{\dagger}_{{\bf q},2}
 + h.c. \right]  ,
\end{eqnarray}
where ${\cal N}$ is the number of sites of one sublattice.
The quasiparticle-magnon interaction
vertex is given by
\begin{equation}
\lambda_{\bf k} = 2 J_{\perp} \left[
2 \lambda_{1}
\left( \cos 2 k_{x} \cos k_{y} + \cos k_{x} \cos 2 k_{y}
\right)
+ \lambda_{2}
 \left( \cos 3 k_{x}  +  \cos 3 k_{y} \right)  \right] .
			\label{lambk}
\end{equation}
The dispersion $ \epsilon_{\bf k} $ is given
by Eq.~(\ref{epsk_1}).

\section{Quasiparticle properties}      \label{sec3}

The quasiparticle dispersion parametrized by
 $\alpha_{1}$
and $\alpha_{2}$(see Table~\ref{hop_vert})
correctly captures all qualitative features found
in the numerical studies of
the $t-J$ model with one
hole.\cite{trug90,dago90,liu92,poil93,poil93_1,preu,dago,putz}
These include (i) the right magnitude of band width
given by $4(\alpha_{1}+2\alpha_{2}) $
(ii) the tight-binding dispersion of the
 same-sublattice hopping type with the
bottom of the band at $(\pi/2, \pi/2)$, and
(iii) the small energy difference between
the quasiparticle energies at
the momenta $(\pi/2, \pi/2)$ and $(\pi, 0)$.
This  small energy difference results from  the closeness
between the values $\alpha_{1}$ and $ 2 \alpha_{2}$. In our
approach,
this is due to the fact that   a quasiparticle
can hop from the site ${\bf r}$ to
${\bf r} + {\bf x} +{\bf y}$
by exchanging spins either between the sites ${\bf r}$
and ${\bf r}+{\bf x}$ or between ${\bf r}$
and ${\bf r}+{\bf y}$. In contrast,
the two spins on the sites  ${\bf r}$
and ${\bf r}+{\bf x}$ must be exchanged in order
to move the quasiparticle from  ${\bf r}$
 to ${\bf r} + 2{\bf x}$
as illustrated in Fig.~\ref{hop1}.
Taking $t=1$ and $J_{\perp}=J_{z}=0.4$ as an example,
our dispersion is given by
$ \epsilon_{\bf k}=  0.387 \cos k_{x} \cos k_{y}
+ 0.110 (\cos 2 k_{x} + \cos 2 k_{y} )$. This is
to be compared with
$ \epsilon_{\bf k}=  0.34 \cos k_{x} \cos k_{y}
+ 0.13 (\cos 2 k_{x} + \cos 2 k_{y} )$
reported in the literature.\cite{dago94,dago}
{}From Table~\ref{hop_vert} and by setting $J_{\perp}=0$,
we notice that the
quasiparticle has a small dispersion even
for the $t-J_{z}$ model and the minimum
of the band is located at ${\bf k}=(0,0)$,
in agreement with the previous
result.\cite{dago90,poil93_1}

The quasiparticle spectral weight is determined by
the overlap between a propagating
bare hole
and the quasiparticle wave function $|{\bf k}\rangle $.
{}From Eqs.~(\ref{qmove_k}) and (\ref{e0r}),
it is simply given by $u_{0}^{2}/N_{\bf k}$.
For different $t/J_{z}$ ratios, the quasiparticle
spectral weight at the bottom
of the band $(\pi/2, \pi/2)$ is listed in
Table~\ref{hop_vert}.
Within our first order degenerate perturbation in
$J_{\perp}$,
the effect of the spin fluctuations on the internal
structure of the quasiparticle is not included.
The consequence is that the quasiparticle spectral weight
only depends on $J_{z}/t$.
In the exact diagonalization study of the isotropic
$t-J$ model, the spectral weight is momentum
dependent. Generally, the spectral weight is
found smaller
for the high-energy states of the quasiparticle dispersion.
This reshuffling of the spectral weight could be the reason
for the systematic underestimation of the spectral weight
in our first order degenerate perturbation.
Physically, one can imagine that the quantum spin
fluctuations smear the energy gap between
the localized string states of the $t-J_{z}$ model
and spread the spectral weight of the
first excited state of the $t-J_{z}$ model
toward the low energy side.
If the nature of the quasiparticle is not  changed
when including the $J_{\perp}$ effect on
the quasiparticle structure, one expects to
improve
the first order degenerate perturbation
by optimizing the coefficients
$u_{l}$ in Eq.~(\ref{e0r}).
Eder and Ohta have achieved this by
constructing a quasiparticle operator
in which the coefficients corresponding to $u_{l}$
are optimized against their exact diagonalization
spectral function.\cite{eder94p}
 Nevertheless, our
first order degenerate perturbation results
give a fairly good estimate of the quasiparticle spectral
weight
at the ground state wave vector $(\pi/2, \pi/2)$ of the
isotropic $t-J$ model.\cite{dago94,poil93,poil93_1}

Besides the effect of the quantum spin fluctuations
on the quasiparticle structure,
the quasiparticle motion also suffers
scattering from the spin wave excitations due to
the interaction terms in  $H_{fb}$.
However, the second order corrections to the quasiparticle
dispersion and spectral weight usually does not
 exceed a few percent
if we use the values $\lambda_{1}$ and $\lambda_{2}$
listed in Table~\ref{hop_vert}.
This is most due to the fact that
 the vertex $\lambda_{\bf k}$
 given by (\ref{lambk}) vanishes
around the bottom of the band.
To the accuracy of applying the
first order degenerate perturbation
to the isotropic $t-J$ model,
these corrections can be neglected.
Strictly speaking, the spin wave excitations can mediate
long range interactions between quasiparticles in the
presence of N\'{e}el order.
However, for the interaction vertex  given by
$\lambda_{\bf k}$
the disorder in real materials
should localize the quasiparticles
before the effect of the long range interactions
becomes significant.
 For completeness, we include the
 expression for the second order self-energy of
 the quasiparticle here. For the one-hole problem
 with $J_{z}=J_{\perp}=J$, we found
 \begin{equation}
 \Sigma({\bf k}, i\omega_{n}) =
 \frac{1}{2{\cal N}} \sum_{{\bf q} \in AFBZ}
 \frac{   \lambda^{2}_{{\bf k}+{\bf q}}
 +  J \lambda_{{\bf k}+{\bf q}} [2 \lambda_{{\bf k}+{\bf q}}
 - \lambda_{\bf k}
 ( \cos q_{x} + \cos q_{y})] / \omega_{\bf q}   }
 {i\omega_{n} - \omega_{\bf q} - \epsilon_{{\bf k}+{\bf q}}
} ,
 \end{equation}
where $\omega_{n}=(2n+1)\pi T$ is  the Matsubara frequency
and
$\omega_{\bf q} = 2 J \sqrt{1 - (\cos q_{x}+\cos
q_{y})^{2}/4}$.
We note that for the one-hole problem the chemical
potential $\mu$ is chosen such that $\epsilon_{\bf k} \geq
0$
for all ${\bf k}$.

\section{Conclusion } \label{sec4}

We have developed a degenerate perturbative treatment
of the $t-J_{z}-J_{\perp}$ model based on the observation
that the quasiparticle has the same nature as a string
state.
Evidence supporting this picture
is abundant from early variational
calculations\cite{trug88,trug90} to the
more recent exact diagonalization
studies\cite{dago90,eder94p}.
A prerequisite for the success of the variational
calculation
is that the elementary excitations can be described by
a hole dressed with strings of overturned spins.
Recently, Eder and Ohta calculated the one-particle spectral
function using a composite operator
 precisely describing a string state $|{\bf r} \rangle$
given by (\ref{e0r})
except with optimized coefficients $u_{l}$.\cite{eder94p}
They found that the quasiparticle peak is greatly enhanced
while the incoherent part is suppressed when the
composite operator is used instead of the
bare hole operator.
This unambiguously  reveals the nature of the quasiparticle
since by progressively improving
the composite operator  one will presumably
 reach a point where only the quasiparticle peak
is left. Thus, an exact quasiparticle operator
can be constructed at least conceptually.

Is the quasiparticle discussed in this paper related to the so-called
spin bag, introduced by Schrieffer et al.\cite{schrieffer88}?
We will postpone the detailed discussion of this question to another
publication.\cite{ganhedegardlee} The original spin bag was introduced
in connection with the Hubbard model, and the bag is a local reduction
of the N\'{e}el order parameter
associated with the bare hole. The size of
the spin bag is reduced when $U$ is increased, whereas the
quasiparticle of the $t-J$ model, discussed here, is increasing in
size, when $J$ is decreased, i.e., when $U$ is increasing. In fact, we
shall argue, that in the limit of large $U$  Hubbard model the
spin bag
will become the bare hole in the $t-J$ model. We shall present
a unified treatment of the Hubbard model in the future
publication.\cite{ganhedegardlee}.

With finite doping, it is known that the long
range antiferromagnetic order is destroyed.
If the quasiparticle is still formed and stable, the
propagation of the
quasiparticle will suffer frustration at the length scale of
the magnetic
correlation length because of the fluctuations of the
local spin directions.
Traveling along a closed path, the quasiparticle will pick
up a Berry phase, resulting in a loss of phase coherence.
These issues will also be
discussed in the future publication.

%\newpage

\acknowledgments
The authors acknowledge useful discussions
with Dung-Hai Lee and
thank D.~Poilblanc for providing the exact
diagonalization results used  for comparison in
Table~\ref{hop_vert}.

	%\end{multicols}

\clearpage

\begin{table}
\caption{Average energies of the string configurations of
different lengths $l$. }
\begin{tabular}{cccc}
 $l$  & $\epsilon_{l}/J_{z}$  & $\epsilon'_{l}/J_{z}$ for
${\bf r}_{s}-{\bf r}=2{\bf x}+{\bf y}$ &
 $\epsilon'_{l}/J_{z}$ for ${\bf r}_{s}-{\bf r}=3{\bf x}$
\\
\tableline
0 & 1  &  3 & 3 \\
1 & 2.5 & 4.5 & 4.5 \\
2 & 3.5 & 5.375 & 5.4583 \\
3 & 4.3889 & 6.0139 & 6.2639 \\
4 & 5.1296 & 6.7454 & 6.963  \\
5 & 5.8704 & 7.412 & 7.6142 \\
6 & 6.5535 & 8.0828 & 8.286 \\
7 & 7.2531 &    &   \\
\end{tabular}
\label{elvalue}
\end{table}

\begin{table}
\caption{ The parameters of effective
Hamiltonian evaluated using the
string state approximation. $Z_{h}({\bf k})$ is
the quasiparticle spectral weight.
The values of $Z_{h}$ in the parentheses
are  the exact diagonalization
results of the $26$ site cluster
at the ground state wave vector ${\bf k}^{*}$
for the $t-J$ model from
Refs.~\protect\cite{poil93,poil93_1}.
Note that,  in Ref.~\protect\cite{poil93_1},
$Z_{h}$ increases with the cluster size
for the $t-J_{z}$ model at  $J_{z}/t=0.3$
while $Z_{h}$ at ${\bf k}^{*}$
decreases with the cluster size for the isotropic
$t-J$ model at $J/t=0.3$.
 }
\begin{tabular}{cccccc}
 $J_{z}/t$ & $Z_{h}(\pi/2,\pi/2)$ &
$\alpha_{1}$  & $\alpha_{2}$ & $\lambda_{1}$ & $\lambda_{2}$
\\
\tableline
0.2    &  0.160(0.213) & $ 0.305 J_{\perp} -0.077 J_{z}$ &
 	$ 0.157 J_{\perp} +0.010 J_{z}$ & 0.348 & 0.124
\\
0.3    &  0.220(0.285) & $0.293 J_{\perp} - 0.047 J_{z}$ &
 	$0.147 J_{\perp} + 0.004 J_{z}$ & 0.297 & 0.101
\\
0.4    &  0.275        & $0.272 J_{\perp} - 0.030 J_{z}$ &
	$0.135 J_{\perp}+0.002 J_{z}$ & 0.253 & 0.083    \\
0.5    &  0.324(0.395) & $ 0.250 J_{\perp} -0.020 J_{z}$ &
	$0.124 J_{\perp} + 0.001 J_{z}$ & 0.218 & 0.070
\\
0.6    &  0.370 & $0.230 J_{\perp} -0.013 J_{z}$ &
	$0.114 J_{\perp}$  & 0.188 & 0.060
\\
\end{tabular}
\label{hop_vert}
\end{table}

\pagebreak

\begin{figure}
\epsfysize=16cm
%%%XXX\epsfbox{fig1.ps}
\caption{ An example with a winding path showing
the nonorthogonality
of the approximate  ground states
$\protect{ |{\bf r}\rangle } $
of the $t-J_{z}$ Hamiltonian.
The configuration is generated
by successively moving the hole
in the  perfect N\'{e}el configuration
  through the sites 012301. But
this is a configuration with a string of length 1
 from  site 2 to 1.
   }
 \label{nonorthog}
\end{figure}

\pagebreak

\begin{figure}
\epsfysize=16cm
%%%XXX\epsfbox{fig2.ps}
\caption{ The upper configuration has a hole separated from
a pair of overturned spins by only one lattice site.
The lower configuration has a hole separated from one
isolated
overturned spin by two lattice sites. Three hops of the hole
in the lower configuration generate
the upper configuration.
   }
 \label{holeflip2}
\end{figure}

\pagebreak

\begin{figure}
\epsfysize=16cm
%%%XXX\epsfbox{fig3.ps}
\caption{ This is a configuration
with a string of length
four
starting from the site 1. Spin
exchange between sites 1
and 2
transforms it to a configuration
with a string of length 2
starting from  site 3.
   }
 \label{hop1}
\end{figure}

\pagebreak

\begin{figure}
\epsfysize=16cm
%%%XXX\epsfbox{fig4.ps}
\caption{ Same configuration as in
Fig.~\protect{\ref{hop1}}.
Spin exchange between sites 2 and 3 transforms it to a
configuration
with a string of length 1  starting from  site 4 and
an isolated overturned spin at site 1.
   }
 \label{vertex}
\end{figure}

\end{document}